\documentstyle[elsart12]{elsart}
\begin{document}
\unitlength 1mm
%
%
\begin{frontmatter}
\title{Continuous control of ionization wave chaos\\
	  by spatially derived feedback signals}
\author{Th.~Mausbach, Th.~Klinger, and A.~Piel}
\address{Institut f\"ur Experimentalphysik\\
	Christian-Albrechts-Universit\"at zu Kiel\\
	Olshausenstra\ss e~40--60, D-24098~Kiel, Germany}
\author{A.~Atipo, Th.~Pierre and G.~Bonhomme}
\address{L.P.M.I., URA~835 du CNRS\\
	Universit{\'e} Henri Poincar{\'e}, BP~239\\
	F-54506 Vand{\oe}uvre-l{\`e}s-Nancy Cedex, France}
\maketitle
\date{\today}
\begin{abstract}
In the positive column of a neon glow discharge, two different types of 
ionization waves occur simultaneously. The low--dimensional chaos arising 
from the nonlinear interaction between the two waves is controlled by a  
continuous feedback technique. The control strategy is derived from the 
time--delayed autosynchronization method. Two spatially displaced points
of observation are used to obtain the control information, using the  
propagation characteristics of the chaotic wave.\\
PACS Numbers: 52.35.-g, 05.45.+b, 52.80.-s
\end{abstract}
\end{frontmatter}
%
%
%
%
Due to their inherent nonlinearity, instabilities in plasmas often develop towards
chaotic dynamics and turbulence \cite{lichtenberg92}. In many practical cases 
this is considered as an undesired situation and there is a particular 
interest to influence the plasma system in order to achieve a stationary 
state (fixed point in the phase space) or a state of regular motion (limit
cycle in the phase space). The most straightforward approach would be to 
change the set of discharge parameters to establish a new non--chaotic state. 
This, of course, may be cumbersome or even impossible. Recent results 
\cite{pierre96} have demonstrated the efficiency of chaos control in 
laboratory plasmas. Moreover, recent computational studies of chaos 
control strategies \cite{satherblom97} offer the possibility of applications
in fusion plasmas. 
In the present communication we present a simple continuous feedback 
technique to control chaotic states in plasmas arising from the nonlinear interaction of
waves. The efficiency of the control scheme is demonstrated experimentally
for ionization wave chaos in the positive column of a glow discharge.

Typically, an infinite number of unstable periodic orbits (UPOs) is embedded
within a chaotic phase space attractor \cite{eckmann85}. This observation motivates the idea 
to achieve the stabilization of selected UPOs by means of small time dependent 
perturbations of an accessible control parameter. Ott, Grebogi, and Yorke 
(further referred to as OGY) have proposed an elegant control strategy based 
on the stabilization of fixed points in the Poincar\'e section \cite{ott90} 
that has achieved a broad field of practical applications \cite{shinbrot93}.
The main obstacle for a universal use of OGY to control chaotic states is the 
necessity for the online determination of the Poincar\'e mapping. 
Therefore, in non--driven (autonomous) chaotic systems the OGY
scheme is limited by computational speed. 
In such situations a control strategy based on information obtained in the 
time domain becomes advantageous. A particular method that uses the 
information of previously recorded dynamics to determine the required control 
information was proposed by Pyragas \cite{pyragas92,pyragas93}. It is referred 
to as time--delay autosynchronization (TDAS) method. If, for simplicity, a 
two-dimensional dynamical system is considered, TDAS is described by the 
autonomous system \cite{pyragas92} 
\begin{equation}
\begin{array}{lll}
	\dot{Y} &=& P(X,Y)+F_\tau(Y)\\ 
	\dot{X} &=& Q(X,Y)\,, 
\end{array}
\end{equation}
where $\{X(t), Y(t)\}$ represents the state of the system. The unperturbed 
phase space flow is given by the functions $P$ and $Q$. The control
signal $F_\tau(Y)$ is obtained by the linear control law 
$F_\tau(Y)=K\cdot[Y(t-\tau)-Y(t)]$
where $\tau$ denotes an appropriate time delay. It is chosen equal 
to the fundamental period of the oscillation signal that corresponds to a 
particular UPO. The constant factor $K$ determines the feedback strength and
$F_\tau$ vanishes if control of the desired periodic orbit is achieved.
The TDAS control method has been applied succesfully to experimental chaotic
systems, for instance lasers \cite{bielawski94}, electronic circuits
\cite{pyragas93,bielawski93} and plasma discharges \cite{pierre96}. It was 
extended later for better performance \cite{socolar94,pyragas95} and 
its mechanism can be understood to some extent now \cite{just97}. 
TDAS and its variants meet technical limitations, too. 
The time--delayed signal $Y(t-\tau)$ is obtained either by delay lines 
\cite{pyragas93,bielawski94,bielawski93} (fast systems) or digitally stored 
data \cite{pierre96} (slow systems). The subsequently discussed control 
scheme may overcome such technical problems. 

Starting with a chaotic state, the wave character of dynamics allows us
to stabilize unstable periodic orbits. Considering the fact that temporally 
periodic states are related to a wavenumber by the dispersion relation, the 
time delay $\tau$ may be replaced by a spatial displacement such that the 
condition ${\omega}/{k}=v_\varphi={\zeta}/{\tau}$ holds,where $\tau$ is 
fixed by the period of the orbit to be stabilized. Here the delay is chosen
as ${\tau}={\Delta}z/v_{\varphi}$ with ${\Delta}z=n\lambda$. The resulting
control law now reads $F_\zeta(Y)=K\cdot [Y(z-\zeta,t)-Y(z,t)]$.
It is easily realized by using a differential amplifier and two spatially
displaced detectors. $K$ is a constant gain factor that may be determined 
experimentally. The control signal $F_\zeta(Y)$ is applied to an accesible
dynamical quantity.

For an experimental demonstration we have chosen the chaotic dynamics of 
ionization waves propagating in the positive column of an ordinary
glow discharge tube. This system is a representative example for
spatially extended dynamics that contain a wavetrain with a large number of
wavelengths. In certain parameter regions of the pressure $p$ and the
discharge current $I_d$, the positive column is either homogenous or different 
types of ionization waves occur \cite{oleson68,pekarek71}. The nonlinear 
dynamics of ionization waves in glow discharges has already been investigated
in great detail. Low--dimensional chaos was first observed in autonomous 
systems with the discharge current as control parameter 
\cite{braun87,braun92}. In the non--autonomous case, where the discharge 
current is modulated by an external periodic driver signal, low--dimensional 
chaotic phase space attractors were studied in detail 
\cite{ohe80,wilke90,albrecht93,weltmann93}. In the latter case, a simplified 
variant of the OGY scheme has already been successfully implemented and UPOs 
up to periodicity 32 could be stabilized \cite{weltmann95}. Note, that 
the success of this 
approach relies on constructing the Poincar\'e section by making use of the 
periodic external driver signal. In the case of an autonomous chaotic system
it was recently demonstrated in a discharge similar to the one under 
investigation that control could be achieved by a slightly modified variant 
of the TDAS scheme \cite{pierre96}. 

%
%
%
%

\begin{figure}[htb]
  \begin{center}
  \begin{picture}(50,50)
   \put(0,48){\special{em:graph three.pcx}}
  \end{picture}
  \end{center}
\caption{Time series of the integral light emission fluctuations 
	corresponding to the $p$--wave (top) and the $s$--wave (bottom). 
	The time series are recorded at $z_p=50\,{\rm mm}$ (close to anode) 
	and $z_s=580\,{\rm mm}$ (close to cathode). The waves are
	mode--locked where the frequency matching condition $f_s=2f_p$ is met 
	in the midsection of the discharge tube (center).}
\label{fig:waves}
\end{figure}
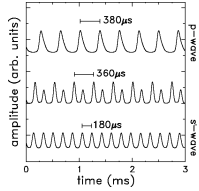

The experimental investigations are performed in a conventional cold-cathode 
glow discharge tube. The discharge length is $l=600\,{\rm mm}$ and the tube 
has a radius of $r=15\,{\rm mm}$. The ionization waves are observed by picking 
up the light emission flux $\Phi (z,t)$ with two movable optical fibres 
connected to fast photodiodes. The spatial resolution is estimated to be 
$\Delta z\leq 1\,{\rm cm}$, which is well below the typical wavelength of 
ionization waves. The discharge is operated at a pressure of 
$p=1.8\,{\rm mbar}$ with neon as filling gas. The discharge current can be 
varied between $I_d=1\dots 50\,{\rm mA}$. In this pressure range a positive 
column forms that extends over 80\% of the discharge length. For the present 
discharge conditions two different types of ionization waves are observed 
simultaneously \cite{krasa74,rutscher64}:
(I) $p$--waves (due to atomic ions) with a frequency of 
$f_p\approx 3\,{\rm kHz}$ and a phase velocity
$v_{\varphi,p}=247\,{\rm m/s}$ directed towards the cathode, and 
(II) $s'$--waves (due to metastables) with a frequency of 
$f_{s'}\approx 5.6\,{\rm kHz}$ and a phase velocity 
$v_{\varphi,s'}=509\,{\rm m/s}$ directed towards the anode. 
Fig.~\ref{fig:waves} shows time series of the waves in a mode--locked state.
Near the anode the $p$--wave dominates, whereas at the cathode a pure $s'$--wave
appears. The frequency matching condition $f_{s'}=2f_p$ is met in the 
midsection of the discharge tube. Since both wave types propagate simultaneously 
in the positive column, they show pronounced nonlinear interaction. 
Consequently a broad variety of dynamical phenomena and low--dimensional chaos
is observed as mentioned above. 

For the practical realization of the control scheme only few conditions have 
to be satisfied. Due to dispersion and the spatial amplification property
of the positive column, the amplitudes of the measured light fluctuations have 
to be equalized, so that the control signal depends only on the phase 
information. The spacing between the two optical fibres has to be exactly 
one wavelength or integer multiples thereof for an optimal control. This 
finding supports the interpretation that the two--point observation acts as
a spatial filter. The  axial position of both fibres, i.e.~the distance from 
the anode is used to adjust the phase of the control signal. 

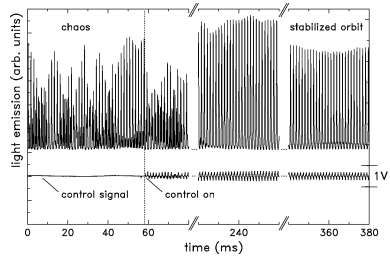
\begin{figure}[htb]
  \begin{center}
  \begin{picture}(100,67)
   \put(0,65){\special{em:graph switch1.pcx}}
  \end{picture}
  \end{center}
\caption{Time series of the integral light emission fluctuations ($Y(z,t)$)
	and the control signal. The control is applied at $t_0=58\,{\rm ms}$
	and the stabilization is achieved within $150\,{\rm ms}$. The 
	modulation degree of the discharge current $I_{mod}/I_{dc}$ does 
	not exceed $5\%$. Smaller $K$-values as well as larger  
	ones lead to a loss of control.}
\label{fig:switch}
\end{figure}
The stabilization of an UPO of a chaotic state arising from the interaction of 
two different ionization waves is demonstrated in Fig.~\ref{fig:switch}. The 
control signal is applied at $t_0=58\,{\rm ms}$.
After approximately $150\,{\rm ms}$ a periodic orbit of periodicity $P=1$ is
fully stabilized as long as the control signal is present. It was not 
possible to stabilize UPOs of higher periodicity. This is, however, an often 
recognized limitation of continuous feedback techniques \cite{pyragas93}. 
The remaining 
modulation of the control signal in Fig.~\ref{fig:switch} is caused by small 
differences in the shape of the fluctuation signals due to the different 
observation points. The control method is nevertheless quite efficient since 
the open loop control would require a much higher amplitude for the suppression 
of chaos \cite{ding94}. 
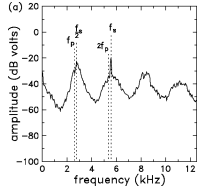
\begin{figure}
  \begin{picture}(60,50)
   \put(20,48){\special{em:graph fft1.pcx}}
   \put(60,50){\special{em:graph bchaos.pcx}}
  \end{picture}
\caption{In (a) the power spectrum of the time series is shown in the chaotic 
	state. Both frequencies ($f_p$ and $f_s$) according to the specific 
	ionization waves are observable. In (b) the reconstructed phase space
	of the corresponding time series is shown. The chaotic attractor 
	($D_2=3.5$) is embedded into a smaller space of three dimensions. 
	The cube axes correspond to a time lag of $\tau=4$ between 
	successive data points. The number of data points for the 
	reconstruction of the phase space vectors is $N=10000$.}
\label{fig:fft1}
\end{figure}
\begin{figure}
  \begin{picture}(60,50)
   \put(20,48){\special{em:graph fft2.pcx}}
   \put(60,50){\special{em:graph bcont.pcx}}
  \end{picture}
\caption{In (a) the power spectrum of the time series is shown in the 
	controlled state. Only $f_s$ is remaining due to the stabilization 
	of both waves. In (b) the reconstructed phase space of the 
	corresponding time series is shown. Due to the reduction of 
	attractor dimensionality ($D_2=1$) the embedding into the 
	three--dimensional space is sufficient. The cube axes correspond to 
	a time lag of $\tau=4$ between successive
	data points. The number of data points for the reconstruction of the
	phase space vectors is $N=10000$.} 
\label{fig:fft2}
\end{figure}
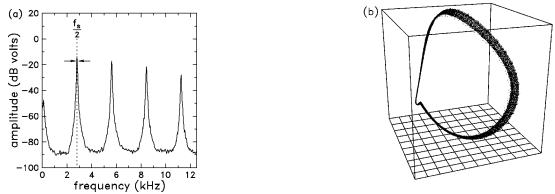
The application of control leads to a dramatic 
decrease ($\approx$70dB) of broad--band components in the power spectrum 
[see Figs.~\ref{fig:fft1}a and \ref{fig:fft2}a]. In the chaotic 
state, the two ionization waves occur as small maxima in a broad spectrum,
whereas in the controlled state only one frequency is established. The 
inspection of the reconstructed phase space diagrams 
[see Figs.~\ref{fig:fft1}b and \ref{fig:fft2}b] illustrates the reduction of 
the attractor's dimensionality. The correlation dimension \cite{grassberger83} 
of the chaotic state has been estimated to be 
$D_2\!=\!3.5\pm 0.3$. 
\begin{table}[htp]
\caption{\label{table}Lyapunov spectra and corresponding dimensionalities 
	 \protect\cite{kaplan79} for the chaotic and the controlled state.}
\begin{center}
\begin{tabular}{lccccc}
 state & $\lambda_1$ & $\lambda_2$ & $\lambda_3$ & $\lambda_4$ & 
 $D_{\rm KY}$\\
 \hline
 chaotic & $0.34\pm 0.05$ & $0.15\pm 0.05$ & $0.06\pm0.08$ & 
 $-0.74\pm0.10$ & $3.74\pm 0.33$\\
 stable & $0.02\pm 0.03$ & $-0.95\pm 0.10$ & -- & -- & $1.02\pm 0.03$\\
\end{tabular}
\end{center}
\end{table}
The estimation of the spectrum of Lyapunov exponents and 
the corresponding Kaplan-Yorke dimension \cite{kaplan79} (Table \ref{table}) 
was done with the computation program of Kruel and Eiswirth \cite{kruel93} 
based on the algorithm of Sano and Sawada \cite{sano85}. 
It shows that two positive 
Lyapunov exponents are dominating the dynamics ($D_2\simeq D_{KY}$). The 
uncertainty is estimated by comparing the results of several calculations. 
After the stabilization, the dimensionality is reduced to $D_2=1.0\pm 0.2$ 
(limit cycle).

%
%
%
%
To summarize, the control of ionization wave chaos in a neon glow 
discharge by a continuous control technique where the feedback signal is 
derived from spatial displacement of detectors has been demonstrated 
experimentally. The control information is easy to determine, the feedback 
scheme is quite simple and even fast dynamical systems can be controlled in 
which digital electronic fails. In contrast to the TDAS--approach, 
no time--delay lines and phase shifting circuits are required. The optimum 
control signal depends sensitively on the precise distance between the optical 
fibres and the absolute $z$-position due to transit time effects and 
differing shapes of the wave in the positive column. 

This control strategy is expected to be more efficient than earlier attempts 
to suppress chaos by a simple open loop control \cite{ding94} and could be 
of major interest in the various chaotic situations frequently observed in 
plasmas. 
\ack
This work was supported by the DFG under Grant No. Pi 185-10. T.M.~Kruel and
M.~Eiswirth are acknowledged for their generous policy in distributing their
programs for the computation of Lyapunov spectra. The French--German 
co--operation has been supported by the HC\&M contract of the E.U.~(CHRXCT 
930356) through the network ``Nonlinear phenomena in microphysics of 
collisionless plasmas. Application to space and laboratory plasmas''.
%
%
%

\end{document}